\documentstyle[12pt]{article}
\def\by#1#2{{\displaystyle {#1}\over \displaystyle {#2}}}
\def\d{{\rm d}}

\begin{document}
\begin{flushright}
IMSc/2002/11/38 \\
25 Nov 2002
\end{flushright}
\begin{center}
{\Large \bf Implications on neutrino oscillation plus decay from recent solar
neutrino data} \\
D. Indumathi, \\
{\it The Institute of Mathematical Sciences, Chennai 600 113, India} \\ [1cm]
\end{center}

\paragraph{Abstract}:
Recently the Sudbury Neutrino Observatory (SNO) has measured both the
rates as well as the day-night asymmetries in the flux of neutrinos from
the Sun in the charged-current (CC), neutral-current (NC) and elastic
scattering (ES) channels. Motivated by the small but negative day-night
asymmetry in the NC channel at SNO (1.2 standard deviations away from
zero), we consider whether such a non-zero asymmetry can be generated
in a framework where active neutrinos can both oscillate and decay,
since the NC day-night asymmetry is zero in a frame-work that includes
pure oscillations of active flavours. We find that the neutral current
day-night asymmetry is strictly zero when we include both oscillations
and decay. The result holds for arbitrary number of active neutrino
species, with some reasonable assumptions on the decay parameters. Hence,
a non-zero asymmetry in the neutral current sector, if it survives,
can only come from oscillations/decay into sterile flavours. We also
examine the allowed region of parameter space coming from a combined
analysis of the total event rates in the charged-current, neutral current
and elastic scattering sectors in a two-flavour framework, and find that
the neutral current rates are most sensitive to (and hence restrict
considerably) the allowed values of the decay parameter involving the
life-time of the heavier mass eigenstate to $\tau_{2_0} > 10^{-4}$ s for
eV mass neutrinos.


\section{Introduction}

There is now very good evidence for neutrino flavour oscillation from
solar neutrino data \cite{homestake,Gallex,SuperK,SNO}. Combined analyses of
these data seem to indicate \cite{Bahcall,analyses} a preference for the
so-called Large Mixing Angle (LMA) solution in the mass-squared
difference--mixing angle ($\delta$-$\theta$) parameter space (in the case
of two-flavour mixing); on inclusion of a third flavour, these results
still apply to the 1--2 mixing angle and corresponding mass-squared
difference ($\delta_{12}$-$\omega$), $\omega \equiv \theta_{12}$, since the
1--3 mixing angle ($\phi \equiv \theta_{13}$) is known to be small
\cite{chooz}.) This LMA solution is sensitive to solar matter effects
on the propagating neutrino in a mildly energy-dependent way. The
Small Mixing Angle (SMA) solution which is also a viable solution
\cite{Bahcall,BIM} to the solar neutrino problem, on the other hand,
has significant energy dependence due to solar matter effects. A detailed
study of the various available data on solar neutrinos in order to isolate
the allowed parameter regions in oscillation parameter space has been
the subject of many rigourous analyses \cite{analyses}. Analyses
of data including decay have also been done for example in
Refs.~\cite{Anjan,Sruba,Beacom}. Detailed analysis of models including
oscillation and decay, applicable to atmospheric and supernova neutrinos
as well, can be found in Ref.~\cite{Ohlsson}. In this paper, we
concentrate on some constraints specifically arising from solar neutrino
data (both from the rates and day-night asymmetries), particularly from
the neutral current (NC) sector, at the Sudbury Neutrino Observatory
(SNO) \cite{SNO}.

Information on Earth matter effects on neutrino mixing and hence flavour
survival and conversion probabilities is available through a study of
the day-night asymmetry, $A^{ND}$, defined as
$$
A^{ND} \equiv 2 \by{{\cal{N}}^N - {\cal{N}}^D}{{\cal{N}}^N + {\cal{N}}^D}~,
$$
where ${\cal{N}}^i$ is the event rate measured during $i = $ Night,
Day. Here data from elastic scattering (ES) of neutrinos on electrons at
Super-K \cite{SuperK} has been augmented by data from SNO \cite{SNO}. The
SNO data includes the day-night asymmetry in ES, in charged current (CC)
$\nu_e\, d$ interaction as well as in the NC $\nu_x \, d$ interaction.

The SNO and Super-K ES data are consistent with each other, within
error-bars. They indicate a small day-night effect. The SNO CC
asymmetry is positive (about 2 $\sigma$ away from zero) while the NC
asymmetry is barely negative (1.2 $\sigma$ away from zero).

It is well-known that oscillations among purely active neutrino flavours
cannot generate a NC day-night asymmetry. In this paper, we examine
whether it is possible to include interactions that generate even a small
(whether positive or negative) day-night asymmetry in the NC sector. We
find that this is impossible even with the inclusion of decay into the
oscillation framework. In other words, if, for example, neutrino decay
occurs through Majoron emission,
$$
\nu_2 \to \overline{\nu}_1 \, \phi~,
$$
even the inclusion of both decay effects in the neutrino sector
as well as of the antineutrinos that are produced in the decay {\it does
not} yield rates that depend on the Earth-matter effects (that is, whether
it is day or night at the time of detection).  Hence the NC day-night
asymmetry remains zero, as in the pure oscillation case. This is true for
arbitrary number of (active) flavours, with some assumptions on the decay
lifetimes.

We then analyse the rates in the NC, CC, and ES channels in a
two-flavour framework, which is justified since the 1--3 mixing angle is
small \cite{chooz}. The parameters are the two mixing parameters
$(\delta_{12}$ and $\omega)$, and a decay parameter $\alpha_2$ that
involves the life-time of the heavier mass eigenstate. It turns out
that the contribution of the antineutrinos to the rates in any channel is
small (depending on the life-time of $\nu_2$). Also, there is not much
variation in the allowed oscillation parameter space for a large range
of the decay parameter $\alpha_2$ in the CC and ES channels. However,
the NC channel is very sensitive to $\alpha_2$ and thereby to the
oscillation parameters. In fact, it turns out that the NC channel
is the most sensitive measurement that constrains the decay
parameter $\alpha_2$.

In summary, the day-night NC asymmetry is strictly zero in a framework
involving both neutrino oscillations and decay (of active neutrino
species). Of course, pure oscillations cannot also generate a day-night
NC asymmetry. The contribution of anti-neutrinos to CC and ES day-night
asymmetry is also small. Hence a clean signature for antineutrinos and
hence for decay cannot come from the observation of such asymmetries:
the best method remains to look for such signals in the {\it rates}
rather than in the asymmetries: this will be signalled by two neutrons
in coincidence with positron in the CC sector at SNO. However, the decay
parameter itself is rather tightly constrained from the NC rates data.

In Section II we calculate the day-night asymmetries in a scenario
that includes both neutrino oscillation and decay, for the NC,
CC and ES channels. We show that the NC asymmetry is zero (leaving
details to the Appendix), with some reasonable assumption on the decay
life-times. Section III contains some numerical results on the allowed
mixing parameter space from analysis of the CC, NC and ES rates at SNO
\cite{SNO} and the corresponding day-night asymmetries.  Section IV
contains a short discussion and conclusions. The Appendix includes some
details of the day-night asymmetry calculations.

\section{Neutrino and antineutrino survival probabilities with decay}

We consider a framework where neutrino flavours $\nu_\alpha$ are mixtures
of the mass eigenstates, $\nu_i$, and the heavier mass eigenstates
$\nu_i$, $i \neq 1$, decay with life time $\tau_i$ ($\nu_1$ is assumed
to be stable). We have,
\begin{equation}
\vert \nu_\alpha \rangle = \sum_i U_{\alpha i} \vert \nu_i \rangle~,
\label{eq:decayfact}
\end{equation}
where the mixing matrix $U_{\alpha i}$ can be expressed in terms of the
mixing angles. We ignore any phase dependence in the mixing matrix. The
time evolution of the heavier mass eigenstates includes a decay term
apart from the phase factor, $\exp(-i E t)$:
$$
\vert\nu_j(t)\rangle = \exp(-i E_j t) \exp [-t/(2\gamma_j \tau_{j_{0}})]~
\vert \nu_j (0) \rangle~,
$$
where $\tau_{j_{0}}$ is the life-time in the neutrino rest frame and
$\gamma_j = E_j/m_j$ is the usual Lorentz factor.

The (electron-type) neutrinos are produced in the core of the Sun. On
their way out, they can interact with solar matter, and then reach the
Earth. Depending on whether it is day or night at the detector, the
neutrinos may further pass through Earth matter. The neutrino survival
(or conversion) probabilities thus depend on the mixing matrix in the
Sun, in the Earth, and the life-time of the heavier eigenstates. We will
begin with the well-known expressions for the day-time probabilities.
Details are included in the Appendix.

\subsection{Day rates}
The neutrino produced in the core at time $t_0$ propagates as a mass
eignestate $\nu_i$ in the Sun until the time $t_R$ when it can undergo
non-adiabatic transitions to another eigenstate, $\nu_j$. It then
continues to propagate outwards, leaving the Sun at time $t_1$ and
reaching the Earth at time $t_2$, where it is detected.  The probability of
a flavour $\alpha$ produced in the core being detected as flavour $\beta$ on
Earth in the day-time can be written in a factorised form \cite{Mohan} as,
\begin{equation}
\langle P_{\alpha\beta}^D \rangle = \sum_j P_{\alpha j}^S P_{j \beta}
f_j^D~,
\label{eq:pabd}
\end{equation}
where the superscript $S$ indicates the dependence on solar parameters.
The braces over $P$ indicate an averaging over the resonance region (or
equivalently, an average over the production region \cite{Mohan}). We
will always assume such an averaging and drop the braces in what
follows. Note that the assumption of such a phase averaging restricts the
neutrino mass-squared difference to be at least $\sim 10^{-8}$ eV${}^2$.

The first term in the expression above represents the probability that
the produced flavour $\nu_\alpha$ will traverse the solar matter and
exit it as the mass eigenstate $\nu_j$. The second term is the (vacuum)
probability of detection of this state as the flavour $\nu_\beta$ in
the detector. The factor $f_j^D$ contains the decay information which
also factors out, provided we make a single assumption about the decay
life-times, as follows.

From Eq.~\ref{eq:decayfact}, we see that, apart from the phase factor
that arises on propagation, the probability amplitude includes decay
terms that involve factors of the type,
\begin{eqnarray} \nonumber
 & \sim & \exp\left[-\int_I^F \by{\d t}{2\tau_j}\right]~, \\ \nonumber
 & = &    \exp\left[-\int_I^F \by{m_j}{2\tau_{0_{j}}} \by{\d t}{E_j}\right]~,
\end{eqnarray}
where $(I, F)$ refer to any of the time intervals $(t_0, t_R)$, $(t_R,
t_1)$, or $(t_1, t_2)$. Now, if the decay rate is very large (life-time
is small compared to the time taken to traverse the Sun-Earth distance),
most of the neutrinos would have decayed by the time they reached the
Earth. This is certainly ruled out by observations. Hence, the relevant
life-times are at least of the order of the Sun-Earth distance. Due to
this, only the time interval $(t_1, t_2)$ (when the neutrino propagates
from the Sun to the Earth) is relevant for decay, and the other decay
factors involving shorter time intervals can be neglected. With this
single assumption, the decay factor $f_j^D$ factorises out of the
probability expression in Eq.~\ref{eq:pabd}. Note that $f^D = 1$ for
the case of pure oscillations as well as the lightest (stable)
neutrino. Detailed expressions for these terms are given in the
Appendix. We note here that $f^D_j$ can be written in terms of
the parameter $\alpha_j$ as $f_j^D = \exp(-\alpha_j/E_j)$, where
$\alpha_j$ involves the average Sun-Earth distance $R$ and a
matter-independent term: $\alpha_j = R m_j/\tau_{j_{0}}$ \cite{Anjan}
where $m_j$ are the (unknown) masses of the heavier eigenstates.

Similarly, the probabilities for an electron neutrino to be detected
as an antineutrino, $P_{\alpha{\overline{\beta}}}$, at the detector
can be computed and are also dependent on the decay parameters
$\alpha_j$.  An equation similar to Eq.~\ref{eq:pabd} holds, with
the matter-dependent mixing angle, $\omega_m$, being redefined with
the opposite sign for the matter dependent term. For details, see the
Appendix.

Finally, the event rate in the day time can be obtained by multiplying the
incident flux, $P_{e\beta} \Phi_e$ with the corresponding cross-section
for detecting $\nu_\beta$ over the detector volume.

\subsection{Night rates}

It turns out that the conversion probabilities at night can also be
written in a factorised form as the product of probabilities. While this
is straightforward for the case of pure oscillations, in the case of
decay, the factorisation again needs the assumption on the decay
life-times used earlier. We have (see the Appendix for details),
\begin{equation}
P_{\alpha\beta}^N = \sum_j P_{\alpha j}^S P_{j \beta}^E f^N_j~,
\label{eq:pabn}
\end{equation}
where $P_{\alpha j}^S$ is the same as for the day probability,
but the second term includes an Earth matter effect that depends
on the path length traversed through the Earth. Furthermore, within
the approximations used, it turns out that $f^N_j = f^D_j$.  Again,
a similar equation (with $A \to -A$ in the matter-dependent terms) holds
for the case when $\nu_\beta$ is an antineutrino. It is important that the
factorised form of the probabilities holds in both the day and night cases,
for particle and antiparticle production; this is a crucial ingredient
in showing that the NC asymmetry is zero.

Number conservation, or unitarity, implies that, for any $j$,
\begin{eqnarray} \nonumber
\sum_\alpha P_{j\alpha} & = & 1~, \\ \nonumber
\sum_\alpha P_{j\alpha}^E & = & 1~.
\end{eqnarray}
The first expression is valid in vacuum, the second in Earth.
Similar expressions hold for the antineutrino case as well. In the
neutral current channel, all flavours contribute equally. The
expression that occurs in the NC night rates calculation below is,
\begin{eqnarray} \nonumber
\sum_\alpha P_{e\alpha}^N & = & \sum_j P_{ej}^S f_j^N \sum_\alpha
                                 P_{j\alpha}^E~, \\
		      & = & \sum_j P_{ej}^S f_j^N~,
\label{eq:sumn}
\end{eqnarray}
where we have used unitarity in the second line. This
is {\em independent} of the Earth-matter effects and hence is the same
as the day rate expression for the same sum that involves the vacuum
probabilities, $P_{j\alpha}$:
\begin{eqnarray} \nonumber
\sum_\alpha P_{e\alpha}^D & = & \sum_j P_{ej}^S f_j^D \sum_\alpha
                                 P_{j\alpha}~, \\
		      & = & \sum_j P_{ej}^S f_j^N~,
\label{eq:sumd}
\end{eqnarray}
where we have used $f_j^D = f_j^N$. This equality also holds for the
sum $\sum_\alpha P_{e\overline{\alpha}}$. This is a key ingredient in
our result.

Trivially, the following sum rule is satisfied:
$$
\sum_\alpha \left[ P_{e\alpha} +
P_{e\overline{\alpha}} \right] = 1~.
$$

\subsection{Day-Night Asymmetries}
We are now ready to write down expressions for the asymmetries
observed at SNO. There are three asymmetries, corresponding to the
CC, NC and ES events. We begin with the neutral current asymmetry. The
cross-sections for the three cases are known, both for incident neutrinos
and antineutrinos \cite{Bahcallb,Kubodera}. It turns out that the
cross-section for neutrinos (or antineutrinos) on deuterium is flavour
independent in the NC channel. This is another crucial ingredient to
our null result on the pure NC asymmetry including decay.

The total number of events in the day-time for the NC case is given by,
\begin{eqnarray} \nonumber 
{\cal N}_{NC}^D & = &  \sum_\alpha \left[ {\cal N}_{\alpha}^D +
{\cal N}_{\overline{\alpha}}^D \right]~, \\ \nonumber
              & = & K_{NC} \Phi_e \sum_\alpha \left[ P_{e\alpha}^D
	            \sigma^\alpha(E_\nu) + P_{e\overline{\alpha}}^D
		    \sigma^{\overline{\alpha}}(E_{\overline{\nu}})
		    \right]~,
\end{eqnarray}
where $P_{e\overline{\alpha}}$ involves both the energy of the produced
$\nu_e$ and that of the decay product, $\overline{\nu}_\alpha$. A similar
expression holds for the night-time rate, with the superscript $D$
being replaced by $N$ everywhere. The overall constant factor $K_{NC}$
includes the detector volume (number of target atoms) as well as the
time over which the measurement is made\footnote{Due to differences
in lengths of day over the year, the day and night data are usually
accumulated over different times; hence, the rates are to be normalised
appropriately when calculating the asymmetry.}.  We now make use of the
equality of the NC cross-sections, $\sigma^\alpha = \sigma$ for all
flavours, $\alpha$, and write,
\begin{eqnarray} \nonumber
{\cal N}_{NC}^D & = & K_{NC} \Phi_e \left[\sigma \sum_\alpha
                      P_{e\alpha}^D + \overline{\sigma}
		      \sum_\alpha P_{e\overline{\alpha}}^D
		    \right]~, \\ \nonumber
{\cal N}_{NC}^N & = & K_{NC} \Phi_e \left[\sigma \sum_\alpha
                      P_{e\alpha}^N + \overline{\sigma}
		      \sum_\alpha P_{e\overline{\alpha}}^N
		    \right]~, \\ 
		& = & K_{NC} \Phi_e \left[\sigma \sum_\alpha
                      P_{e\alpha}^D + \overline{\sigma}
		      \sum_\alpha P_{e\overline{\alpha}}^D
		    \right]~.
\end{eqnarray}
We see from Eqs.~\ref{eq:sumn} and \ref{eq:sumd} that the day and night
rates (when accumulated over the same period) are the same. Hence the
NC day-night asymmetry,
$$
A_{NC}^{N\!D} = 0~,
$$
in a frame-work including both oscillation and decay of (active)
neutrino flavours.  This result does not hold if the decay occurs to a
sterile flavour, since neutrinos are then ``lost'' and number conservation
(the unitarity argument used above) does not hold. Finally, note that in
deriving this result, no assumptions have been made on the details of
the matter-dependence either in the Sun or in the Earth.

For completeness, we list the other asymmetries in the Appendix. The CC
asymmetry involves only the electron neutrino interactions while
the ES asymmetry includes the muon neutrino as well (and of course the
antiparticles). We now move on to a numerical analysis of the data.

\section{Numerical Calculations for rates and asymmetries}

In this section, we compute the event rates in the CC, NC, and ES channels
in an oscillation frame-work including decay. For simplicity, we work in a
2-flavour scenario so the parameters involved are the mixing parameters
$\delta_{12}$ and $\omega$ and the decay parameter $\alpha_2$. In
a 3-flavour frame-work, apart from additional mixing angles, $\phi =
\theta_{13}$ and $\psi = \theta_{23}$, and the corresponding mass-squared
differences, there is another decay parameter, $\alpha_3$. However, there
is very little sensitivity to the 1--3 mixing angle $\phi$ \cite{BIM}
which is known to be small, and no dependence at all of solar neutrino
rates on the 2--3 angle $\psi$.  Furthermore, in the limit of small
$U_{e3}$ (and hence small $\phi$), which is driven by the CHOOZ result
\cite{chooz}, it was shown in Ref.~\cite{Anjan} that the $\alpha_3$
dependence drops out of the expressions for the day-time rates so the
decay is still governed only by the second-heaviest eigenstate through
$\alpha_2$. In short, the 2-flavour analysis is sufficiently sensitive to
most of the features of the model; we shall therefore present numerical
results within such a frame-work.

The detailed expressions for the rates in a two-flavour frame-work are
listed in the Appendix. We shall use these formulae to explicitly
compute the rates in the CC, NC, and ES channels. We use a chi-squared
minimisation program to best-fit the free parameters and obtain the
allowed regions in the parameter-space. We also compute, for
completeness, the day-night asymmetry in these channels corresponding
to the best-fit values of the parameters.

\subsection{Allowed parameter region from rates analysis}

As stated before, we use the cross-sections given by Ref.~\cite{Kubodera}
for the neutrino-deuteron CC and NC cross-sections and that given by
Ref.~\cite{Bahcallb} for the ES cross-sections.  We assume a nearly
massless majoron so that the antineutrino produced by decay has half
the energy of the initial neutrino.

We use the fiducial volume of 0.77 ktons of heavy water and the zenith
angle dependence for 128.5 ``days'' of day data and 177.9 ``days'' of
night data as given by SNO \cite{SNOdata}. We use the normalisation of
NC events (with kinetic energy of detected electron $T_e > 5$ MeV) as
indicated by SNO \cite{SNOdata}. We also use the resolution function for
the CC, ES and NC case as given by SNO. For the theoretical predictions,
we use the standard solar model of Bahcall, Pinsonneault and Basu
\cite{bp2001}. We use the Boron neutrino flux and ignore the hep
contribution.  Note that there are about 20\% errors on the Boron flux
of neutrinos.

We use a simple one-slab model of the Earth with an average density of 5
gm/cc. Due to the latitude location of SNO, it detects neutrinos that
have passed through the (outer) core of the Earth roughly 8.7\% of the
time. (SNO does not see any solar neutrinos that pass through the inner
core of the Earth). By assuming a one-slab model that only incorporates
the mantle, we have effectively neglected the contribution from this
fraction of events. We use the 2-flavour expressions for the
probabilities as given in the Appendix; the parameters to be determined
by comparison with the SNO data are the mass squared difference
$\delta_{12}$, the mixing angle $\omega$ and the decay parameter
$\alpha_2$.

Before we begin the actual numerical calculation, we note that the NC
event rate is amenable to some approximations. Since the antiparticle
cross-section is small (the cross-section depends non-linearly on the
energy and the average energy of the antiparticles is half that of
the particles), the contribution from the antineutrino sector can be
neglected. (This is also true in the CC case). We have,
\begin{eqnarray} \nonumber
{\cal{N}} & \propto & \Phi \left[ (P_{e1}^S + P_{e2}^S \exp(-\alpha_2/E_2))
\sigma + P_{e2}^S (1- \exp(-\alpha_2/E_2)) \overline{\sigma}\right]~,
\\
 & = & \Phi \left[ \sigma - P_{e2}^S (1- \exp(-\alpha_2/E_2))
           \left\{\sigma - \overline{\sigma} \right\} \right]~, \\
 & \sim &  \Phi \sigma \left[1 - P_{e2}^S (1- \exp(-\alpha_2/E_2))
 \right]~.
\end{eqnarray}
Note that the NC events are independent of mixing in the absence of
decay. Hence, when $\alpha_2 = 0$, the second term is zero and the NC
rate is independent of the mixing parameters. In fact, it is proportional
to $\Phi \sigma$, which is the total rate in the absence of neutrino
oscillation/decay. As decay is turned on (so that $\alpha_2 \neq 0$),
the NC rate begins to depend on the mixing parameters through $P_{e2}^S$
($ = 1 - P_{e1}^S$ for two-flavours). Since the NC rates are well within
that predicted by the standard solar model, to counteract the effect of
a non-zero $\alpha_2$, the parameter space in mixing gets squeezed in
order to minimise $P_{e2}^S$ (or to maximize $P_{e1}^S$) so that the
second term remains small. Since it is not possible to squeeze
$P_{e1}^S$ beyond a point (as it is bounded by unity), larger values of
$\alpha_2$ get ruled out by the SNO result that the NC event rates are
within 1 $\sigma$ of the standard solar model \cite{SNO}. Hence it turns
out that the NC channel is the best suited to constrain the decay
parameter. Finally, such a squeezed solution will be a bad fit for the CC
and ES rates, since here the rates are depleted compared to
expectations. Hence a combined analysis of the CC, NC, and ES data will
severely constrain $\alpha_2$.

With this heuristic discussion, we proceed to calculate the rates from
different channels at SNO. We have performed a combined chi-squared
analysis of the total rates at SNO. The CC and NC rates are strongly
anticorrelated while the CC/NC and ES rates are mildly anticorrelated;
however, we have not included this correlation in our analysis.

While SNO CC data favours a LOW solution (low mass squared
difference, large mixing angle) we know that when combined with data
from other experiments such as GALLEX \cite{Gallex}, this solution is
ruled out \cite{BIM}. Hence we concentrate on solutions, including decay,
around the ``pure oscillation'' LMA and SMA solutions. When $\alpha_2 =
0$ the solutions are determined by the CC and ES rates alone. In fact this
turns out to have the lowest $\chi^2$ value when $\alpha_2$ is turned on.
The best-fit values to the parameters for different values of $\alpha_2$
are shown in Table~\ref{tab:chisq}. It is seen that $\alpha_2 = 5$ MeV
is disallowed at 2 $\sigma$.

\begin{table}
\centering
\begin{tabular}{|l|l|l|l|l|} \hline
Set & $\alpha_2$ MeV & $\delta_{12}$ eV${}^2$   & $\omega^\circ$ & $\chi^2$ \\
\hline
LMA &  0         & 2.7 10${}^{-5}$ & 33.5     & 0.18     \\
    &  1         & 2.7 10${}^{-5}$ & 36.1     & 0.55     \\
    &  5         & 9.2 10${}^{-5}$ & 11.7     & 2.43     \\ \hline
SMA &  0         & 6.4 10${}^{-6}$ &  2.2     & 0.10     \\
    &  1         & 1.9 10${}^{-6}$ &  4.0     & 0.30     \\
    &  5         & 5.8 10${}^{-6}$ &  2.2     & 3.28     \\ \hline
\end{tabular}
\caption{Best fit solutions for fixed $\alpha_2$ from a combined
analysis of CC NC, and ES rates at SNO \cite{SNO} and the corresponding
$\chi^2$ per degree of freedom. Best fit points corresponding to the
so-called LMA and SMA regions are shown. The entire allowed parameter
space is shown in Fig.~\ref{fig:totcontour}.}
\label{tab:chisq}
\end{table}

\begin{figure}[htp]
\vskip 9truecm
\includegraphics{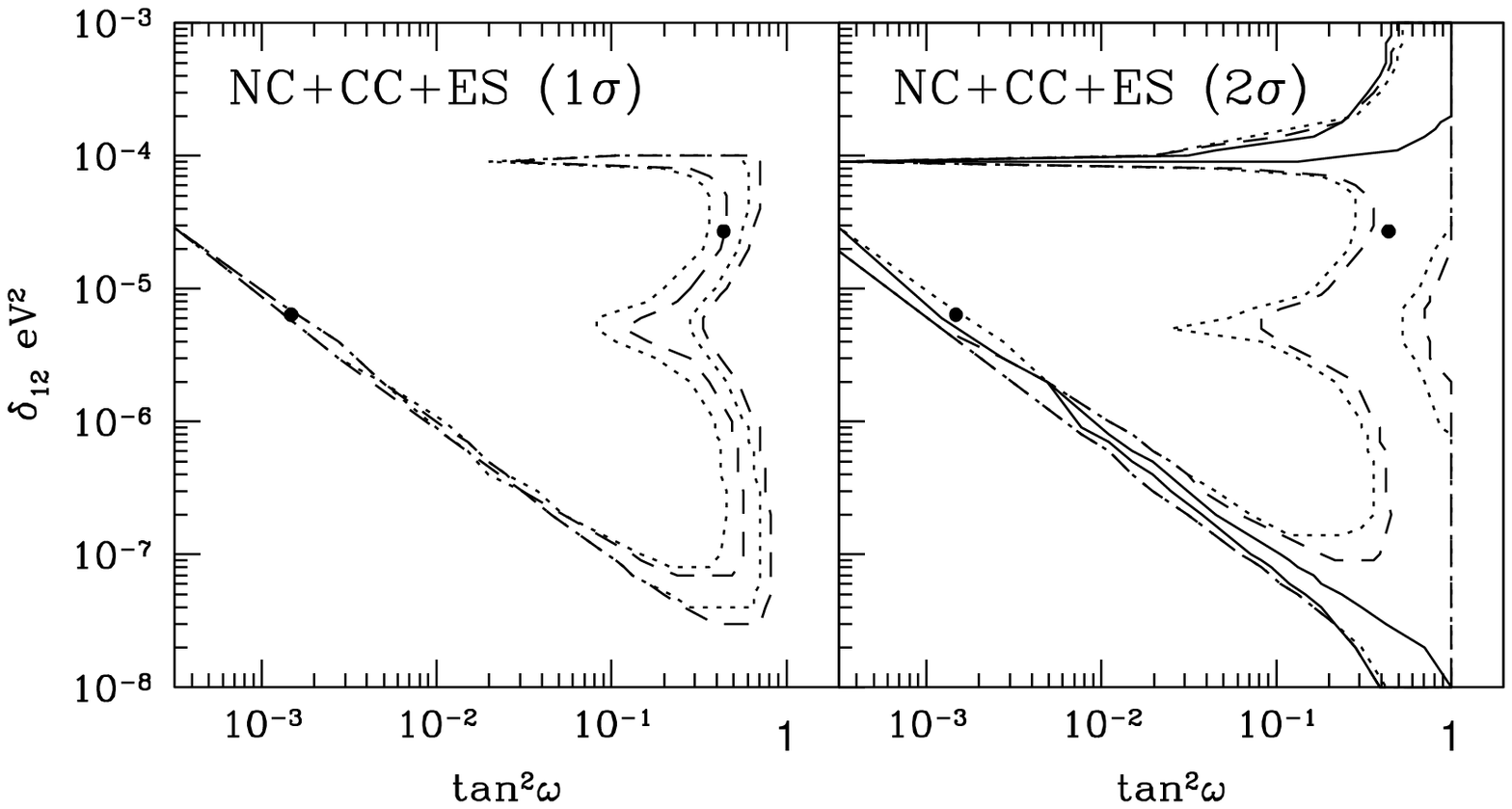}
\caption{1 $\sigma$ and 2 $\sigma$ allowed regions in
($\delta_{12}$--$\tan^2\omega$) space from a combined analysis of
CC, NC and ES rates at SNO \cite{SNO} for fixed values of the decay
parameter $\alpha_2 = 0, 1, 5$ MeV (dotted, dashed and solid lines
respectively). Notice that there are no 1 $\sigma$ allowed regions for
$\alpha_2 = 5$ MeV. The best-fit values (that correspond to $\alpha_2 =
0$) of the parameters in the LMA and SMA region as given in
Table~\ref{tab:chisq} are shown as solid dots
in the figure.}
\label{fig:totcontour} 
\end{figure}

The 1 $\sigma$ and 2 $\sigma$ allowed parameter space from fitting
the combined CC, NC and ES rates in $(\delta_{12}, \tan^2\omega)$ for
$\alpha_2 = 0, 1, 5$ MeV are shown in Fig.~\ref{fig:totcontour}. There
is very little difference between the allowed regions for $\alpha_2 = 0,
1$ MeV. However, there are no solutions at 1 $\sigma$ for $\alpha_2 =
5$ MeV. This is a much stronger result than the limit $\alpha_2 < 18$ MeV
obtained by considering only the CC and ES data \cite{Anjan}. The
2 $\sigma$ allowed regions are shown on the right hand side of
Fig.~\ref{fig:totcontour}. We see that the ``usual'' LMA region for
$\alpha_2 = 5$ MeV is disallowed by the pinching effect on the NC rates
described earlier.

\begin{figure}[htp]
\vskip 9truecm
\includegraphics{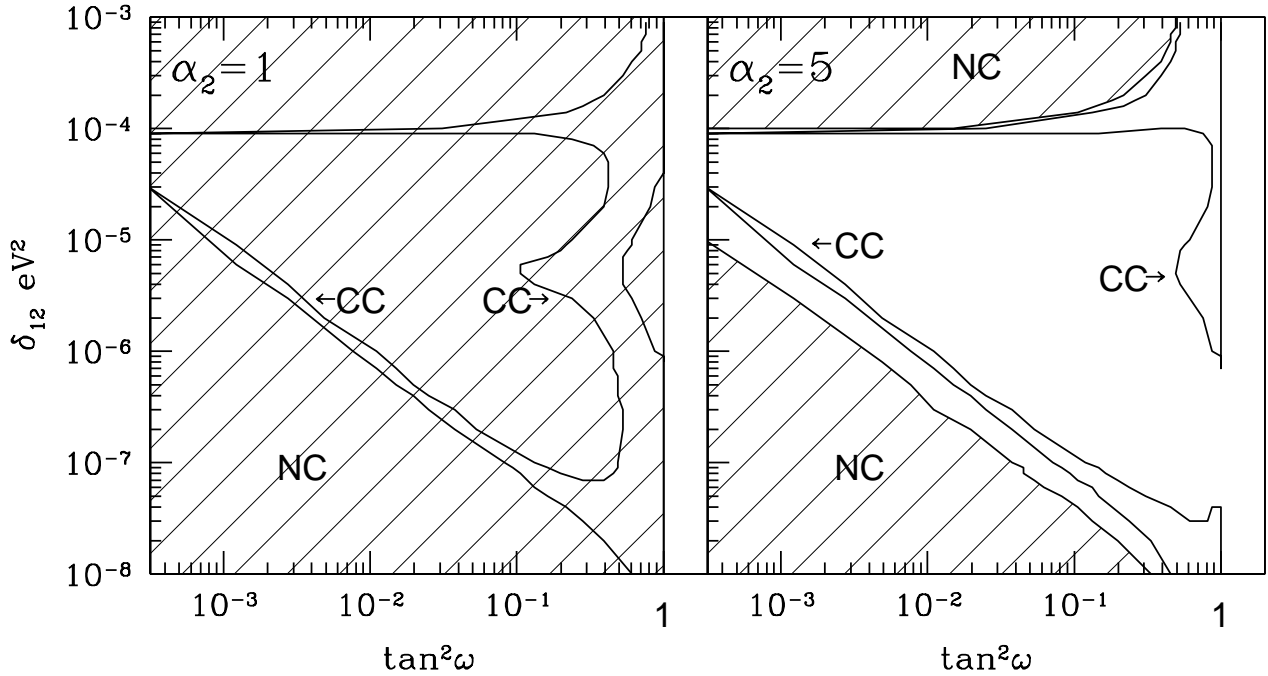}
\caption{1 $\sigma$ allowed regions in
($\delta_{12}$--$\tan^2\omega$) space from individual CC and NC rates
at SNO (calculated with respect to the combined minimum as listed in
Table~\ref{tab:chisq}) for different values of the decay parameter
$\alpha_2$. The NC allowed regions are shaded for contrast. Notice that
there is no overlap between CC and NC allowed regions for $\alpha_2 =
5$ MeV. The allowed regions from ES rates almost completely overlap the
CC ones and hence are not indicated separately.}
\label{fig:contour} 
\end{figure}

This result is easy to understand if we examine the allowed regions
from the individual channels. We find that the 1 $\sigma$ allowed
parameter space in mixing for a given $\alpha_2$ from CC data and ES
data overlap almost completely, while the allowed region from NC data
is very different. The allowed regions coming from individual analyses
of NC and CC rates are shown in Fig.~\ref{fig:contour}.

We know that the NC rate is independent of mixing parameters for
$\alpha_2 = 0$. Even when $\alpha_2 = 1$ MeV, the entire parameter space
is allowed at 1 $\sigma$ level as can be seen from the left hand side
of Fig.~\ref{fig:contour}. For $\alpha_2 = 5$ MeV, there is no overlap
region between the allowed regions from CC/ES and NC data; hence there
are no 1 $\sigma$ allowed regions for such large values of $\alpha_2$.
In the LMA region, for instance, the allowed region from the NC channel
shifts towards smaller $\omega$ in an attempt to maximise $P_{e1}^S$,
as discussed earlier. This results in eliminating the region in parameter
space that is preferred by the CC rate.

Note that we have fitted the total rates and not the energy-dependent
spectrum. Inclusion of larger amounts of decay will distort the spectrum
significantly and so it seems as if analysis of the spectrum may limit
the decay parameters better than the rates. However, as $\alpha_2$
becomes large, there is no allowed parameter space and hence no
observable spectral distortion; a rates analysis is therefore unlikely
to further constrain this result. This is borne out by the consistency
of our results with the detailed spectral analysis, including decay,
found in Ref.~\cite{Sruba}, where a limit $\alpha_2 < 5.7$ MeV is
obtained.

\subsection{The day-night asymmetries}

We calculate the day-night asymmetries for CC and ES channels using the
best-fit values of the parameters for different values of $\alpha_2$
as given in Table~\ref{tab:chisq}. We find that the CC asymmetry is
roughly 0.2 and positive for allowed $\alpha_2$ in the LMA region of
mixing parameters, while the ES asymmetry is slightly smaller and also
positive. The asymmetries are much smaller in the SMA case. When $\alpha_2$
exceeds 4 MeV or so, the asymmetries are nearly zero.

\begin{figure}[htp]
\vskip 15truecm
\includegraphics{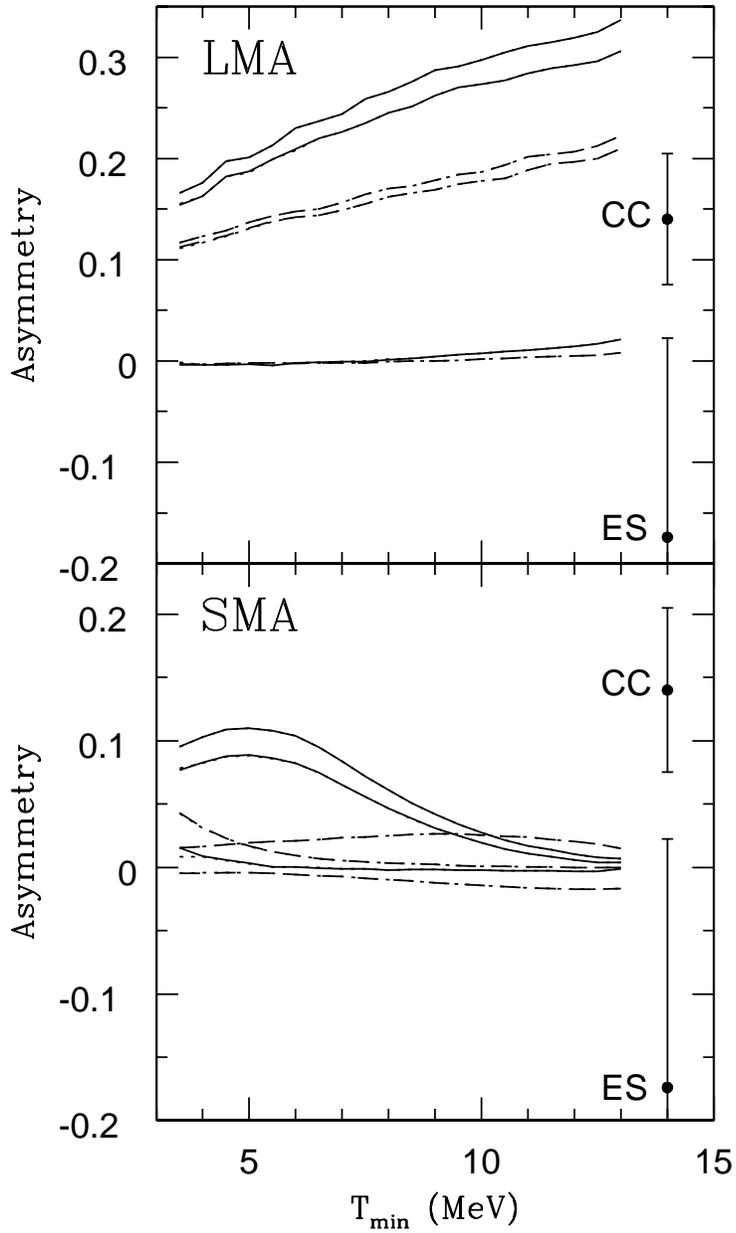}
\caption{The day-night asymmetries for CC and ES (solid and dashed
lines) rates for the best fit parameters for different values
of $\alpha_2$ as a function of the kinetic energy of the observed
electron. The lines correspond to $\alpha_2 = 0, 1, 5$ MeV with decreasing
magnitude of asymmetry. The energy-averaged asymmetries measured by SNO
are shown on the right hand side of the figure, for comparison.}
\label{fig:asymm} 
\end{figure}

The corresponding day-night asymmetries are shown as a function of the
observed kinetic energy of the electron in Fig.~\ref{fig:asymm}.  There is
a negligible contribution from the antiparticles, which contribute only
at low energies, $T < 5$ MeV. The energy-averaged values of the SNO CC
and ES asymmetries are shown alongside for comparison.  Most of the events
correspond to lower energies; hence the average asymmetry is closer to
the value at the lower energy end. There is no significant constraint on the
decay parameter coming from the day-night asymmetries; this of course
may change with the availability of more data from SNO.

\section{Discussion and Conclusions}

We have shown that the neutral current day-night asymmetry is zero,
in a frame-work of neutrino mixing that includes oscillations
plus decay, with a reasonable assumption on the size of the neutrino
life-time. We discussed the effect of decay on the neutral-current,
charged-current and elastic scattering rates in a two-flavour
frame-work. In all cases, it turns out that the contribution from
antiparticles is small in the region of allowed decay parameter space.
This is because of small cross-sections owing to small (roughly half)
energies of antineutrinos compared to those of neutrinos. Hence, even if
there are large decay rates, very few events will be seen. We find that
the neutral-current rates constrain the decay parameter quite
stringently. On comparing with SNO data for the total rates in the CC,
NC, and ES channels, we get $\alpha_2 < 5$ MeV, which translates to a
bound on the life-time of the heavier mass eigenstate of $\tau_{2_{0}}
> 10^{-4}$ s for eV mass neutrinos (equivalently, this is the bound
on $\tau_{2_0}/m$ for an unknown neutrino mass $m$ in eV). This is very
close to the bound that already been obtained in a detailed analysis of
solar neutrino data in Ref.~\cite{Sruba}. This is much
stronger than the bound $\alpha_2 < 18$ MeV obtained in Ref.~\cite{Anjan}
using only the charged-current SNO data; in fact, it is the limiting
bound that can be obtained, given the Sun-Earth distance \cite{Beacom}.
It is interesting that this limit is obtained only on including the
neutral-current data.

Here we have mainly highlighted the dual role of the NC SNO data:
it is most constraining of the decay parameter as far as total {\em
event rates} are concerned; however, the NC {\em day-night asymmetry}
is completely oblivious to the parameters of either mixing or decay. An
in-depth analysis of available solar neutrino data, in a 2- and 3-flavour
frame-work including decay, has been done in Ref.~\cite{Sruba}, where
detailed energy correlations as well as rates from other experiments
such as the Chlorine and Gallium experiments \cite{homestake,Gallex},
have been analysed along with the energy-dependent spectra and day-night
asymmetries from Super-K and SNO \cite{SuperK,SNO}. Our results are
consistent with this detailed analysis.

Given the allowed range of $\alpha_2$ coming from the rates analyses,
we see that the corresponding day-night asymmetries are not very
sensitive to the mixing parameters or indeed to the decay parameter.
Hence, the best way to put limits on, or observe neutrino decay, will
be by a direct measurement of the rates and not the asymmetries. A
distinctive signature of an ``electron'' detected in coincidence with two
neutrons will signal the capture of an antineutrino. This occurs through
the direct observation via the CC interaction in heavy water at SNO:
\begin{equation}
\overline{\nu}_e \, d \to e^+ n \, n ~. 
\end{equation}
Two neutrons must be seen in coincidence with the positron. If
the neutrons are not observed, then this process is indistinguishable from
the dominant one of
$$
\nu_e \, d \to e^- p \, p ~,
$$
since water Cerenkov detectors cannot distinguish electrons from
positrons. Furthermore, both the $\nu_e$ and the $\overline{\nu}_e$
interactions have precisely the same angular dependence (which is $2:1$
backward-forward asymmetric) so that angular distribution cannot be used
to separate the two signals. Finally, even if the antineutrino energy
is half that of the decaying neutrino, SNO will still be able to detect
them through the neutron coincidence measurement. It is expected that
there will soon be a substantial improvement in the neutron detection
efficiency (with the inclusion of salt in the SNO detector); it may
then be possible to detect these channels with good efficiency.

Note here that while the interaction of (electron) antineutrinos on
protons in water is very large (about 100 times the cross-section of
neutrinos in water at solar neutrino energies), Super-Kamiokande will
not see these antineutrinos since their angular distribution is roughly
isotropic and Super-K cuts out non-forward events.

Finally, a remark about the NC day-night asymmetry measured by SNO.
After all, this was the motivating factor for this calculation.
Currently, this asymmetry is barely negative (about 1.2$\sigma$ away
from zero). If, with improved statistics, this asymmetry persists, it
will be difficult to find a source for the effect. It is possible then
that there are sterile neutrinos involved in the oscillation frame-work;
however, the extent of conversion to steriles is limited by the total NC
event rate (which is already consistent with the BP2000 standard solar
model prediction) which admittedly has large theoretical errors (of
about 20\%). Another possible source for a non-zero NC asymmetry could
be that the NC cross-sections for $\nu_e$ and $\nu_\mu$, $\nu_\tau$
scattering on deuteron are not the same.

So, with the improved detection of neutrons, and the accumulation of more
data, SNO will be in a good position to set strong bounds on neutrino
decay and oscillations into sterile flavours.

\vspace{0.3cm}

\noindent{\sf Acknowledgements}: I thank M.V.N. Murthy and
G.~Rajasekaran for many discussions and a critical reading of the
manuscript.

\section*{Appendix}
\setcounter{equation}{0}
We derive the neutrino conversion probabilities in a frame-work that
includes both oscillation and decay.  A flavour $\vert \nu_\alpha\rangle$
that is produced in the core of the sun (actually, an electron type
neutrino), can be expressed in terms of the mass eigenstates $\vert
\nu_i\rangle$, as
$$
\vert \nu_\alpha \rangle = \sum_i U_{\alpha i}^S \vert \nu_i^S \rangle~,
$$
where $U_{\alpha i}$ is the usual mixing matrix that can be expressed in
terms of a mixing angle $\omega$ in the case of two-flavours, or three
angles ($\omega$, $\phi$, $\psi$) for the three-flavour case (we ignore
any CP violating phases).  The superscript $S$ indicates that these are
the mass eigenstates in the Sun. The effect of decay is to modify the
phase factor that determines the time-evolution of the mass eigenstates
so that the time evolution is given by,
$$
\vert \nu_i \rangle_t = \exp [ -i E_i t ] \, \exp [ - t/(2\tau_i) ]
\vert \nu_i \rangle_0~.
$$
The lifetime can be expressed in terms of that in the rest frame:
$\tau_i = \gamma_i \tau_{0_{i}}$, $\gamma_i = E_i/m_i$. 

The neutrino, produced at a time $t_0$ in the core of the Sun,
propagates adiabatically upto the resonance point $t_R$, where it can
undergo non-adiabatic conversion to a different mass eigenstate, and
then continues to propagate adiabatically to the edge of the Sun which
it reaches at $t_1$. Then, depending on whether it is night or day,
the neutrino propagates upto the Earth and gets detected at time $t_2$,
or traverses a further distance $L$ through the Earth and gets detected
at time $t_3$.

Hence, the probability of a flavour $\alpha$ being detected as a
flavour $\beta$ on Earth in the day-time is given by
$$
P^D_{\alpha\beta} = \sum_{ij} \vert U_{\beta j} \vert^2
                    \vert U_{\alpha i}^S \vert^2 
                    \vert M_{j i}^S \vert^2 f_j^D~,
$$
where $U$ is the vacuum mixing matrix, and we have averaged over the
resonance region \cite{Mohan}. The decay factor $f_j^D$ will be defined
below. Here $M_{ji}^S$ represents the non-adiabatic transition probability
amplitude from $i \to j$ so that $\vert M_{ji}^S \vert^2$ is the usual
Landau Zener transition probability \cite{LZ} in the Sun.

As has been pointed out earlier \cite{Mohan}, it is possible to express
the probability as a factor that is solar-matter dependent, times a
solar-matter independent factor:
\begin{equation}
P^D_{\alpha\beta} = \sum_{j} P_{\alpha j}^S P_{j\beta} f_j^D~,
\end{equation}
where $$
P_{\alpha j}^S = \sum_{i} \vert U_{\alpha i}^S \vert^2 
                    \vert M_{ji}^S \vert^2 ~,
$$
and 
$$
P_{j \beta} = \vert U_{\beta j} \vert^2 ~.
$$
As explained in the main part of the text, the decay factor $f_j^D$ in
the expression for the probability amplitude involves a product of terms:
\begin{eqnarray} \nonumber
 & \sim & \exp \left[-\int_I^F \by{\d t}{2\tau_i} \right]~, \\ \nonumber
 & = &    \exp\left[-\int_I^F \by{m_i}{2\tau_{0_{i}}} \by{\d t}{E_i}
            \right]~,
\end{eqnarray}
with one such term coming from every propagation from $t_I$ to $t_F$ of
a neutrino $\nu_i$.
In general, we can assume the modification of the energy eigenvalue due
to the matter-dependent effects is small, so that the integral above
reduces to $\sim (t_F - t_I) [m_i/(2\tau_{0_{i}}E_i)]$~. If the time
interval of propagation is short, then the numerator is small compared
to the life-time unless the latter is itself very small. In the
latter case, we would expect most of the produced neutrinos to have
decayed before reaching the Earth, but this is clearly not true. We
therefore assume that the life-time of the heavier mass eigenstate is at
least the Sun-Earth distance (divided by the speed of light). Then the
decay exponentials are all small and can be dropped, except the terms
containing the time interval $(t_F - t_I) = (t_2 - t_1)$ which is the
Sun-Earth separation. The relevant integral contains the {\it vacuum}
energy eigenvalue and hence is anyway matter-independent. The decay term
that appears in the expressions for the probability is then the square
of the above, with no cross terms, for the energy eigenvalue $E_j$
corresponding to $\nu_j$ that propagates in vacuum from the Sun to the Earth,
$$
f_j^D = \exp (- \alpha_j /E_j )~,
$$
where we have used $\alpha_j = R m_j/\tau_{0_{j}}$ \cite{Anjan} where
$R$ is the Sun-Earth distance.

A similar formula holds at night-time, except that the vacuum
probability $P_{j\beta}$ is replaced by an earth-matter dependent one:
\begin{equation}
P^N_{\alpha\beta} = \sum_{j} P_{\alpha j}^S P^E_{j\beta} f_j^N~,
\label{eq:night}
\end{equation}
with the solar probability the same as before, and with
$$
P_{j \beta}^E = \sum_{kk'} U_{\beta k}^{*E} U_{\beta k'}^{E}
                           M_{kj}^{E} M_{k'j}^{*E} \exp\left[ -i
			   \int_{t_2}^{t_3}(E_k^E - E_{k'}^E) \d
			   t\right]~.
$$
Here $M_{kj}^E = \langle \nu_k^E \vert \nu_j \rangle$~ is the transition
amplitude from $j \to k$, where $\nu_k^E$ is the mass eigenstate {\em
just} inside the Earth. We have assumed only one non-adiabatic
transition occurs, when the neutrino encounters the discontinuity in
density at the Earth's surface. However, it is straightforward to
generalise this to an arbitrary number of discontinuous jumps in Earth's
density, using the so-called slab model of the Earth. Details may be
found in Ref.~\cite{Mohan}. The decay factor is the same as in the
day-time case since the matter effects (whether in Earth or in the Sun)
due to decay are small due to the small distance of propagation in
these. So the vacuum contribution is the only one that remains here as
well. So we have
\begin{equation}
f_j^N = f_j^D~.
\end{equation}
Finally, the event rate can be obtained by multiplying the
incident flux, $P_{e\beta} \Phi_e$ with the corresponding cross-section
for detecting $\nu_\beta$ over the detector volume.

Similar formulae hold for antineutrinos produced during decay, with all
matter-dependent terms being redefined with $A \to -A$, where $A$ is the
matter dependent term in the Sun or Earth. It is important that the
factorised form of the probabilities holds in both the day and night cases,
for particle and antiparticle production; this is a crucial ingredient
in showing that the NC asymmetry is zero.

Number conservation, or unitarity, implies that,
$j$,
\begin{eqnarray} \nonumber
\sum_\alpha P_{j\alpha} & = & 1~, \\ \nonumber
\sum_\alpha P_{j\alpha}^E & = & 1~.
\end{eqnarray}
The first expression is valid in vacuum, the second in Earth. Note that
the second equation is valid even when Eq.~\ref{eq:night} for the
night-time probability is replaced by a more complicated one using a
many-slab model of the Earth.

Similar expressions hold for the antineutrino case as well. In the
neutral current channel, all flavours contribute equally. The
expression that occurs in the NC night rates calculation is,
\begin{eqnarray} \nonumber
\sum_\alpha P_{e\alpha}^N & = & \sum_j P_{ej}^S f_j^N \sum_\alpha
                                 P_{j\alpha}^E~, \\
		      & = & \sum_j P_{ej}^S f_j^N~,
\end{eqnarray}
where we have used unitarity in the second line. This
is {\em independent} of the Earth-matter effects and hence is the same
as the day rate expression for the same sum that involves the vacuum
probabilities, $P_{j\alpha}$:
\begin{eqnarray} \nonumber
\sum_\alpha P_{e\alpha}^D & = & \sum_j P_{ej}^S f_j^D \sum_\alpha
                                 P_{j\alpha}~, \\
		      & = & \sum_j P_{ej}^S f_j^N~,
\end{eqnarray}
where we have used $f_j^D = f_j^N$. This equality also holds for the
sum $\sum_\alpha P_{e\overline{\alpha}}$. From this it is clear that the
NC rates are the same during the day or night for arbitrary numbers of
active neutrino species; hence the NC day-night asymmetry is zero as
explained in the main part of the text. If the decay is into a sterile
flavour, of course, this is no longer true.

We now specialise to the two-flavour case where we list the detailed
expressions for the probabilities needed for the computation of the
rates and hence the asymmetries in the different channels at SNO.
Then with the choice $m_2 > m_1$, $\nu_1$ is stable so that $\tau_2$ is
the only relevant decay parameter. We have,
$$
f_2^D = f_2^N = \exp [-\alpha_2/E_2]~.
$$
We now list the various probabilities. Since the Earth's density is much
smaller than that in the solar core, there is only a small modification
of the energy due to Earth-matter effects. Then we can define
\begin{equation}
X = (t_3 - t_2) \delta_{12}^E /(2E)~,
\end{equation}
where $\delta_{12}^E$ is the mass-squared difference in (Earth) matter:
$$
\delta_{12}^E = \delta_{12} \cos(2(\omega - \omega_e)) - A_E \cos 2
\omega_e~,
$$
where the Earth matter dependent term is, $A_E = 2 \sqrt 2 G_F N_e^E E$
with $N_e^E$ the electron number density in the Earth
and $(t_3-t_2) = 2 R_E \cos\theta_z$ and the matter mixing angle is
$\omega_e$. Here $2 R_E$ is the diameter of the
Earth ($= 5.1 R_E(m) \hbox{MeV/eV}^2$) and $\theta_z$ is the zenith
angle of propagation of the neutrino with respect to the instantaneous
direction of the Sun.

\noindent {\sf The Day-time probabilities}: In the two-flavour case,
$P_{e2} = 1 - P_{e1}$; furthermore, we have, 
\begin{eqnarray} \nonumber
P_{ee}^D & = &  c^2 P_{e1}^S + s^2 P_{e2}^S \exp(-\alpha_2/E_2)~, \\
\nonumber
P_{e\mu}^D & = &  s^2 P_{e1}^S + c^2 P_{e2}^S \exp(-\alpha_2/E_2)~, \\
\nonumber
P_{e\overline{e}}^D & = &  c^2 P_{e2}^S [1 - \exp(-\alpha_2/E_2)]~, \\
P_{e\overline{\mu}}^D & = &  s^2 P_{e2}^S [1 - \exp(-\alpha_2/E_2)]~.
\end{eqnarray}
Here $c$ and $s$ refer to $\cos\omega$ and $\sin\omega$ respectively.
Defining the Landau-Zener probability $P_{LZ}$ \cite{LZ} for transition
between the $\nu_1$ and $\nu_2$ eigenstates (note that the third mass
eigenstate, even if included, does not participate in non-adiabatic
transitions in the Sun, provided the relevant mass squared difference
is that which solves the so-called atmospheric neutrino problem), we have
\begin{equation}
P_{e1}^S = s_m^2 P_{LZ} + c_m^2 (1 - P_{LZ})~,
\end{equation}
where $s_m$ and $c_m$ refer to $\sin\omega_m$ and $\cos\omega_m$ in the
solar matter, given by
\begin{equation}
\cos 2 \omega_m = \by{\delta_{12}\cos 2 \omega  - A}{\sqrt{
(\delta_{12}^2 \sin^2 2\omega + (\delta_{12} \cos 2\omega - A)^2)}}~.
\end{equation}
The above formula holds for antineutrinos with $A \to -A$ and for
neutrinos and antineutrinos in Earth matter as well, with $A \to A_E$ so
that $\omega_m(\pm A) \to \omega_e(\pm A_E)$. 

Note that
\begin{eqnarray} \nonumber
P_{ee}^D + P_{e\mu}^D & = &  P_{e1}^S + P_{e2}^S \exp(-\alpha_2/E_2)~,
\\
P_{e\overline{e}}^D + P_{e\overline{\mu}}^D & = &  P_{e2}^S (1 -
           \exp(-\alpha_2/E_2))~,
\end{eqnarray}
so that the total conversion probability into neutrinos and
antineutrinos is conserved. When $\alpha_2 = 0$ we recover the usual
formulae for pure oscillation case.

\noindent {\sf Night-time probabilities}: Here we need to compute, in
addition, the Earth-matter effects. We have
\begin{eqnarray} \nonumber
P_{1e}^E & = &  1 - 2 c_e^2 s_e^2 - s^2 \cos^2 2 \omega_e + 2 c s c_e s_e
\cos 2\omega_e  \\ \nonumber
 & &   + 2 c_e s_e (c_e s_e \cos 2\omega - c s \cos 2 \omega_e) \cos X~,
 \\ \nonumber
P_{2e}^E & = &  1 - 2 c_e^2 s_e^2 - c^2 \cos^2 2 \omega_e - 2 c s c_e s_e
\cos 2\omega_e  \\ \nonumber
 & &   - 2 c_e s_e (c_e s_e \cos 2\omega - c s \cos 2 \omega_e) \cos X~,
 \\ \nonumber
 P_{1\mu}^E & = & P_{2e}^E~, \\
 P_{2\mu}^E & = & P_{1e}^E~,
\end{eqnarray}
where $X$ has been defined above and $\omega_e$ is the Earth-matter
dependent mixing angle. Notice that, along with the usual number
conservation, $P_{1e}^E + P_{1\mu}^E =1$, we have also $P_{1e}^E
+ P_{2e}^E = 1$~. Substituting in the transition probabilities,
$P_{\alpha\beta}^N$, this gives us,
\begin{eqnarray} \nonumber
P_{ee}^N + P_{e\mu}^N & = &  P_{e1}^S (P_{1e}^E + P_{1\mu}^E) + 
P_{e2}^S (P_{2e}^E + P_{2\mu}^E) \exp(-\alpha_2/E_2) \\
 & = &  P_{e1}^S + (1 - P_{e1}^S) \exp(-\alpha_2/E_2)~.
\end{eqnarray}
Notice that the sum is independent of Earth matter effects and is
exactly equal to the ``Day'' formula; this was already shown in the
general case of arbitrary number of flavours.

For the detection of antiparticles at night, we have similar expressions
to those for the neutrinos, with $\omega_e(A_E) \to \omega_e(-A_E)$:
\begin{eqnarray} \nonumber
P_{e\overline{e}}^N + P_{e\overline{\mu}}^N & = &  P_{e\overline{1}}^S
(P_{1\overline{e}}^E + P_{1\overline{\mu}}^E)~, \\ \nonumber
 & = &  (1 - P_{e1}^S)(1 -\exp(-\alpha_2/E_2))~.
\end{eqnarray}
Again, the total probability is conserved and is independent of the
Earth-matter effects.

\noindent{\sf The event rates}: With these probabilities, we now define
the rates for CC, NC and ES events at SNO. We begin with the neutral
current channel. The event rate at SNO in the NC channel is
given by
$$
{\cal{N}}^{NC} = 
{\cal{N}}^{e} + 
{\cal{N}}^{\mu} + 
{\cal{N}}^{\overline{e}} + 
{\cal{N}}^{\overline{\mu}}  ~.
$$
Hence, 
\begin{eqnarray} \nonumber
{\cal{N}}^{NC}_{Day} & = & \Phi \left[ P_{ee}^D \sigma^e + 
                     P_{e\mu}^D \sigma^\mu + 
                     P_{e\overline{e}}^D \sigma^{\overline{e}} + 
                     P_{e\overline{\mu}}^D \sigma^{\overline{\mu}}
		     \right]~, \\
		     & = & \Phi \left[ \sigma - P_{e2}^S
		     (1-\exp( - \alpha_2/E_2)) \left( \sigma -
		     \overline{\sigma} \right) \right]~; \\ \nonumber
{\cal{N}}^{NC}_{Night} & = & \Phi \left[ P_{ee}^N \sigma^e + 
                     P_{e\mu}^N \sigma^\mu + 
                     P_{e\overline{e}}^N \sigma^{\overline{e}} + 
                     P_{e\overline{\mu}}^N \sigma^{\overline{\mu}}
		     \right]~, \\ \nonumber
		     & = & \Phi \left[ \sigma - P_{e2}^S
		     (1-\exp( - \alpha_2/E_2)) \left( \sigma -
		     \overline{\sigma} \right) \right]~, \\
		     & = & {\cal{N}}^{NC}_{Day}~.
\end{eqnarray}
These expressions leads us to the general result that we had
shown earlier: that the NC day-night asymmetry is zero.
Here $\sigma$ ($\overline{\sigma}$) is the flavour-independent neutral
current cross-section of neutrinos (antineutrinos) on deuteron.  For
calculating the cross-section and hence the rates, we make the simple
assumption that in the decay, the antiparticle carries away about half
the energy of the parent neutrino.  Hence, although the NC cross-sections
for neutrinos and antineutrinos are roughly the same at the same energy,
the contribution from antineutrinos is very small since the cross-section
rises exponentially with energy. Indeed, for most values of $\alpha_2$
the bulk of the contribution to the {\it rates} is still from neutrinos.

Similar expressions can be written for the CC and ES rates. In the
former, only the $\nu_e$ (or $\overline{\nu}_e$) contribute. We have,
\begin{eqnarray} \nonumber
{\cal{N}}^{CC}_{Day} & = & \Phi \left[ P_{ee}^D \sigma_{CC}^e + 
                     P_{e\overline{e}}^D \sigma_{CC}^{\overline{e}} 
		     \right]~, \\
{\cal{N}}^{CC}_{Night} & = & \Phi \left[ P_{ee}^N \sigma_{CC}^e + 
                     P_{e\overline{e}}^N \sigma_{CC}^{\overline{e}} 
		     \right]~,
\end{eqnarray}
where the probabilities involved have been listed earlier and the
cross-sections are taken from Ref.~\cite{Kubodera}. There is no
simplification of the expressions although the anti-neutrino
contribution is small here as well. 

The ES rates involves all particles and antiparticles, but with
flavour-dependent cross-sections as given in Ref.~\cite{Bahcallb}. We
have,
\begin{eqnarray} \nonumber
{\cal{N}}^{ES}_{Day} & = & \Phi \left[ P_{ee}^D \sigma_{ES}^e + 
                     P_{e\mu}^D \sigma_{ES}^\mu + 
                     P_{e\overline{e}}^D \sigma_{ES}^{\overline{e}} + 
                     P_{e\overline{\mu}}^D \sigma_{ES}^{\overline{\mu}}
		     \right]~, \\
{\cal{N}}^{ES}_{Night} & = & \Phi \left[ P_{ee}^N \sigma_{ES}^e + 
                     P_{e\mu}^N \sigma_{ES}^\mu + 
                     P_{e\overline{e}}^N \sigma_{ES}^{\overline{e}} + 
                     P_{e\overline{\mu}}^N \sigma_{ES}^{\overline{\mu}}
		     \right]~.
\end{eqnarray}
The cross-section for elastic scattering of the $\nu_e$ is about 6 times
larger than than of $\nu_\mu$ (and even larger than the antineutrino
cross-sections), although the values are energy dependent. 

\noindent{\sf The day-night asymmetries}:
The CC and ES asymmetries are not zero; the former depends only on
$P_{ee}$ and $P_{e\overline{e}}$. These are different for the day and
night-time. Hence, the CC asymmetry is non-zero by an amount
proportional to 
\begin{equation}
{\cal{N}}^{CC}_{Night} - {\cal{N}}^{CC}_{Day} = \Phi \sum_i \left[
(P_{ei}^N - P_{ei}^D) \sigma_{CC}^i \right]~,
\end{equation}
where $i = e, \overline{e}$.  It is straightforward to similarly write
an expression for the ES asymmetry, which has both charged-current and
neutral-current contributions:
\begin{equation}
{\cal{N}}^{ES}_{Night} - {\cal{N}}^{ES}_{Day} = \Phi \sum_i \left[
(P_{ei}^N - P_{ei}^D) \sigma_{ES}^i \right]~,
\end{equation}
where the summation includes contributions from electrons and muons (and
their antiparticles as well).

\newpage

\end{document}